\documentclass[preprint,showpacs,showkeys,aps,prb]{revtex4}
\usepackage[english]{babel}
\usepackage[dvips]{graphicx}

\begin{document}

\title{Microscopic and Macroscopic Rayleigh-B\'enard Flows : \\
Continuum and Particle Simulations, Turbulence,              \\
Fluctuations, Time Reversibility, and Lyapunov Instability }

\author{
Wm. G. Hoover and Carol G. Hoover                            \\
Ruby Valley Research Institute                               \\
Highway Contract 60, Box 601                                 \\
Ruby Valley, Nevada 89833                                    \\
}

\date{\today}

\pacs{45, 45.20.Jj, 47.10.Df}

\keywords{Rayleigh-B\'enard flows, macroscopic flows, microscopic flows, Lyapunov instability, chaos}

\vspace{0.1cm}

\begin{abstract}
We discuss the irreversibility, nonlocality, and fluctuations, as well as the Lyapunov and hydrodynamic
instabilities  characterizing atomistic, smooth-particle, and finite-difference solutions of the two-dimensional
Rayleigh-B\'enard problem.  To speed up the numerical analysis we control the time-dependence of the
Rayleigh number, ${\cal R}(t)$ , so as to include many distinct flow morphologies in a single simulation.
The relatively simple nature of these computational algorithms and the richness of the results they can yield
make such studies and their interpretation particularly well suited to graduate-level research.
\end{abstract}

\maketitle

\section{Introduction}

``Understanding Turbulence'' is an enduring catch phrase and has been a potential funding source since the early
days of computers.  There is no shortage of reviews ranging from short sketches\cite{b1,b2,b3} to scholarly
studies\cite{b4,b5,b6}.  The vast research literature takes in earth, air, fire, and water as well as the
weather, the sun, aircraft design, and small-box chaos.  Spectra and power-law relations abound.  Mostly
the working fluid is incompressible and often its motion is described as a superposition of modes or
vortices.  Two- and three-dimensional turbulence behave differently, with the flow of energy away from or toward
smaller length scales in these two cases\cite{b4}.  ``Enstrophy'', the squared vorticity [ squared rotation rate ],
is ``conserved'' in two-dimensional incompressible flow\cite{b4}. 

Despite all this information there appears to be more to learn.  How many vortices should we expect to see?
What is the Lyapunov spectrum like?  How localized are the vectors corresponding to the exponents?  The simple
nature of the underlying model, a conducting viscous fluid, the complexity of the flows that result, and the
multitude of computational schemes, all provide opportunities for imaginative approaches and analyses. We
recommend their study and describe our own explorations of what seems to us the simplest problem involving
turbulence, Rayleigh-B\'enard flow\cite{b1,b2,b3,b4,b5,b6,b7,b8,b9,b10,b11}.

The classical Rayleigh-B\'enard problem describes the convective behavior of a compressible, heat conducting,
viscous fluid in the presence of gravity and a temperature gradient.  Here we suppose that the fluid is
confined by a stationary square $L \times L$ box with fixed boundary temperatures.  Despite these simplest
possible of boundary conditions, even in two space dimensions this problem provides interesting {\it internal}
flows of mass, momentum, and energy.    The heat driving these flows enters and exits along the boundaries.
Most of it comes in at the bottom and flows out at the top.  There are two competing mechanisms for the heat flow
from bottom to top.  The simpler of the two is conduction, described by Fourier's law, $Q = -\kappa \nabla T$ .
But {\it mechanical} ( or ``convective'' ) heat flow is possible too and comes to dominate conduction as the
flow begins to move, and continues to grow as the flow eventually becomes turbulent.

Thermal expansion near the bottom of the box provides the buoyancy necessary to carry the hot fluid upward.
Cooling and compression near the top encourages downward flow.  These vertical driving forces due to
temperature and gravity are balanced by the dissipative effects of heat conduction and viscosity which lead
to macroscopic entropy production.  The dimensionless ratios of these effects, the Rayleigh Number ${\cal R}$
and, to a lesser extent, the Prandtl Number ${\cal P}$ [ which we set equal to unity in our work here ] :
$$
{\cal R} \equiv g(\partial \ln V/\partial T)_P\Delta TL^3/(\nu D) \ ; \ {\cal P} = (\nu/D) \ ,
$$
control the overall flow.  The Nusselt number ${\cal N}$ completes the list of dimensionless flow variables.
It is an observable rather than an input.  ${\cal N}$ is simply the ratio of the ( time-averaged, if
necessary ) vertical heat flux to the prediction of Fourier's law :
$$
{\cal N} = (L Q_y/\kappa \Delta T) \  .
$$
With our thermostated sidewalls the definition of the Nusselt Number is somewhat arbitrary.  {\it Entropy
production} is a more appropriate measure of our flows' separation from equilibrium, though we will not
discuss those interesting results here for lack of space.

The convective flow patterns characterizing Rayleigh-B\'enard flow can be stationary, periodic in time,
or chaotic.  It is often possible to observe qualitatively different solutions -- different numbers of
convective rolls for instance -- for the {\it same} external boundary conditions\cite{b11}.  And at very ``high''
Rayleigh numbers [ on the order of a half million or more ], chaotic flows {\it never repeat}.  Chaotic
solutions describe at least two distinct regimes of turbulence\cite{b1,b2}, called ``soft'' and ``hard'',
and distinguished by the form of their fluctuations, Gaussian or exponential respectively\cite{b1,b4,b5}.
The time scales associated with eddy rotation vary from seconds in the laboratory to \ae ons inside the
earth and sun. The richness of Rayleigh-B\'enard flow patterns, even or especially in two dimensions,
together with their illustration of the fundamentals of fluid mechanics, instability theory, nonlinear
dynamics, and irreversible thermodynamics, makes these problems an ideal introduction to the use of
numerical methods in computational fluid dynamics\cite{b2,b8,b9}.

We choose to study here the simplest possible constitutive model, an ideal gas ,
$$
PV = NkT = E = Nme \longrightarrow (\partial \ln V/\partial T)_P = (1/T) \ ;
\ (S/k) = \int_0^L\int_0^L \rho \ln (T/\rho)dxdy \ . 
$$
For simplicity we set Boltzmann's constant $k$, the particle mass $m$, and the overall mass density
$\rho$ equal to unity.  We choose the hot and cold temperatures equal to 1.5 and 0.5 so that
$\Delta T \simeq T$.  Finally, we choose the gravitational constant $g$ so as to give a constant-density
solution of the equation of motion in the quiescent purely conducting case :
$$
-(dP/dy) = \rho g = -\rho (kdT/dy) \rightarrow \{ \ g = (k\Delta T/mL) \equiv (1/L) \
\rightarrow \rho \simeq 1 \ , \ {\rm a \ constant} \ \} \ .
$$  
With these simplifications a square $[ \ N  \equiv L\times L \ ]$-cell system with a Rayleigh number
${\cal R}$ and a Prandtl Number ${\cal P}$ is achieved by choosing the two constitutive properties,
kinematic viscosity $\nu$ and thermal diffusivity $D$ , to satisfy the two definitions :
$$
(\nu/D) \equiv {\cal P} \ ;  \ {\cal R} \equiv (L/\nu)(L/D) \ .
$$

Unlike experimentalists we computational scientists are not limited to physical materials, dimensions,
or boundary conditions. We have the undoubted luxury that our transport coefficients ( as well as the
gravitational acceleration and even the box size ) can {\it all} be time dependent if we like.  In the
simulations reported here we typically use time-dependent transport coefficients, chosen so that the
Rayleigh number increases or decreases [ to check for hysteresis ] linearly with time.  In this way a
whole {\it range} of Rayleigh numbers, with varying roll numbers, kinetic energies, and Lyapunov exponents
( {\it if} the increase is carried out sufficiently slowly ), can all be obtained with a single simulation.

For sufficiently large values of the Rayleigh number ( 4960 or more for the static fixed-temperature
boundary conditions used here\cite{b7} ) one or more viscous conducting rolls form and evolve with time.  With
the convective heat flow directed upward, and on the average balanced by the gravitational forces acting
downward, either stationary, periodic, or chaotic flows can be achieved.  Figure 1 shows how a simple
single vortex can be used to construct initial conditions with one or more vortices.

\begin{figure}
\vspace{1 cm}
\includegraphics[width=1.4 in,angle= -90]{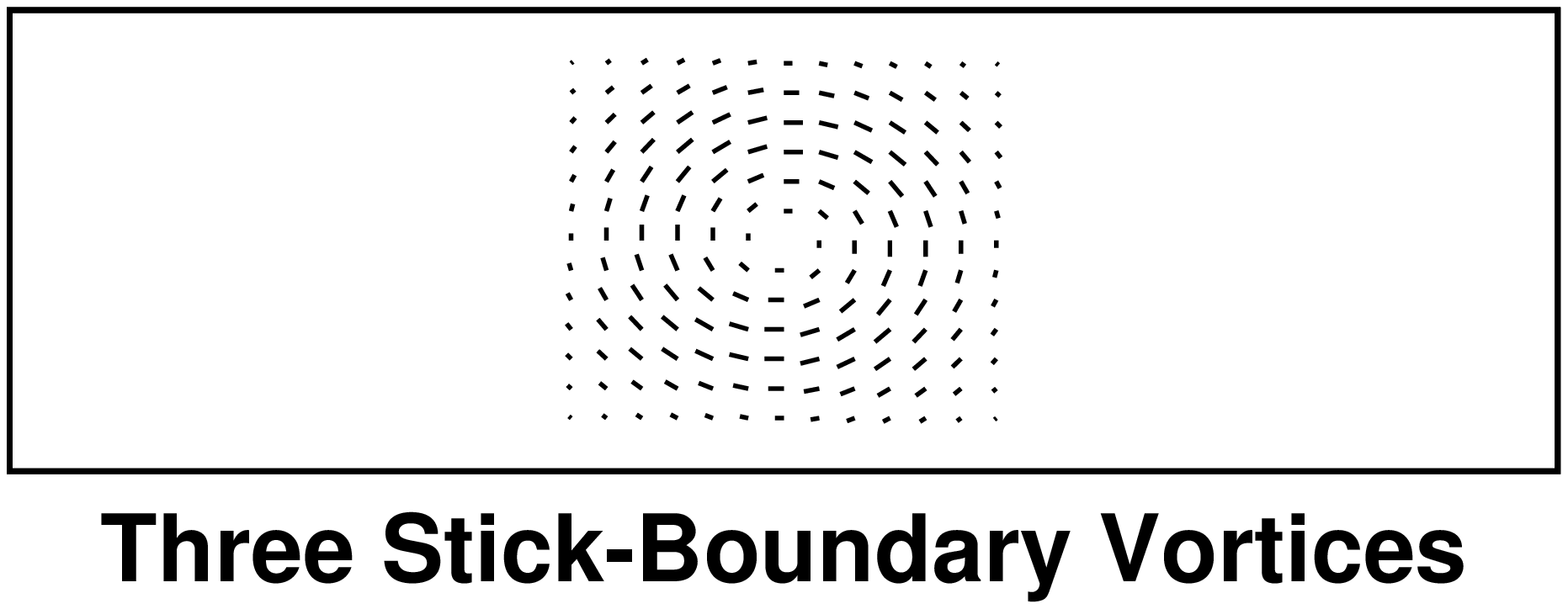}                   
\vspace{5 mm}
\caption{ 
An $L \times L$ square with an idealized roll using the ``stick'' boundary condition
[ zero velocity at the wall ] :
$                                                                                                                       
u_x \propto + \sin(y) \cos(x/2) \ ; \ u_y \propto -\sin(x) \cos(y/2) \ ; \ -\pi \ \leq \ x,y \ \leq \ +\pi \ ,
$
is shown at the center.  These data are also reflected, both to the right and to the left, to make an array
with three rolls.  This illustrates a handy initialization technique to use in the search for stable multiroll
solutions.
}                                                                                                                    
\end{figure}     

Since the 1980s nominally steady-state solutions for such flows have been computed with three distinct
methods: microscopic molecular dynamics together with particle-based and grid-based macroscopic simulation
methods\cite{b8,b9,b10,b11}.  The Smooth-{\it Particle} Applied Mechanics Method ( SPAM ) offers a welcome bridge
between the microscopic and macroscopic approaches\cite{b8}.  In SPAM the dynamics of macroscopic particles
is governed by motion equations including the macroscopic {\it irreversible} constitutive laws. But the
form of those motion equations mimics that of the microscopic motion equations.  In both cases the accelerations are
based on summed-up contributions from neighboring pairs of particles.  SPAM calculations can also be thought
of as a finite-difference algorithm on an irregular grid.

The resulting macroscopic flow patterns exhibit interesting solution changes as the Rayleigh number increases
above the critical value of 4960 .  The {\it positions} of the rolls' centers can exhibit both periodic and
chaotic motion.  Lyapunov exponents characterize the growth of the instabilities leading to chaotic
motion\cite{b4,b7}.  For continuum simulations with thousands of degrees of freedom the simplest calculation of
the instabilities involves only the largest exponent.  There is evidence that the ``spectrum'' of Lyapunov
exponents is roughly linear [ and with a negative sum, due to the dissipative nature of continuum flows ]\cite{b4}.

It turns out that the relative stability of particular flows depends upon the initial conditions.  No
known variational principle ( like maximum entropy, or minimum entropy production ) predicts {\it which} of the
several flows is stable\cite{b11}.  The various ``principles'' based on energy or entropy can be evaluated for
these simulations.  Intercomparisons of the three simulation methods can shed light on the dissipation described
by the Second Law of Thermodynamics and the differing time reversibilities of the microscopic and macroscopic
techniques.

In Section II we summarize the continuum physics of fluid flow problems: mass, momentum, and energy conservation
are always required.  Shear flows and heat flows can result.  In Section III we outline three numerical solution
techniques and display some typical results.  In Section IV we present our conclusions and suggest
research directions useful for students.

\section{Continuum Mechanics and Rayleigh-B\'enard Flow}

A physical description of any continuum flow necessarily obeys the conservation laws for mass, momentum, and energy :
$$
\dot \rho   = -\rho \nabla \cdot u \ ; \
\rho \dot u = -\nabla \cdot P + \rho g\ ; \
\rho \dot e = -\nabla u:P - \nabla \cdot Q \ .
$$
The simplest {\it derivation} of these three laws uses an {\it Eulerian} coordinate system fixed in space.  The
summed-up fluxes of each conserved quantity through the surfaces of each of the $L^2$ computational cells, plus
the internal gravitational contributions give the time rates of change in the cells.  In the Rayleigh-B\'enard
problem the boundary source terms introduce and extract energy at the bottom, along the sides, and at the top,
while the gravitational momentum density source $\rho g$ acts throughout the volume . The fluid's constitutive
properties -- the pressure tensor $P$ and the heat flux vector $Q$  -- are computed from the local state variables
$\{ \ \rho,u,e \ \}$ and their gradients.  $P$ and $Q$ are the momentum and energy fluxes in a coordinate system
``comoving'' with the local velocity $u(r,t)$ .

It is easy to solve the continuum flow laws by converting them to sets of ordinary differential equations.  These
latter equations incorporate the linear phenomenological constitutive relations pioneered by Newton and Fourier,
expressing pressure in terms of the symmetrized velocity gradient, and the heat flux vector in terms of
the temperature gradient :
$$
P = (P_{\rm eq} - \lambda \nabla \cdot u)I - \eta[ \ \nabla u + \nabla u^t \ ] \ ; \ Q = -\kappa \nabla T \ .
$$
$\kappa$ is the thermal conductivity.  We set the bulk viscosity equal to zero ( appropriate for an ideal gas )
by setting $\lambda + \eta = 0$ so that the pressure tensor has the following form :
$$
P_{xx} = P_{\rm eq} -\eta [ \ (\partial u_x/\partial x) - (\partial u_y/\partial y) \ ] \ ; 
$$
$$
P_{yy} = P_{\rm eq} -\eta [ \ (\partial u_y/\partial y) - (\partial u_x/\partial x) \ ] \ ; 
$$
$$
P_{xy} = -\eta [ \ (\partial u_y/\partial x) + (\partial u_x/\partial y) \ ]  \ .
$$
With these constitutive relations specified we have a well-posed continuum problem ready to solve.

\begin{figure}
\vspace{1 cm}
\includegraphics[height=12cm,width=6cm,angle= -90]{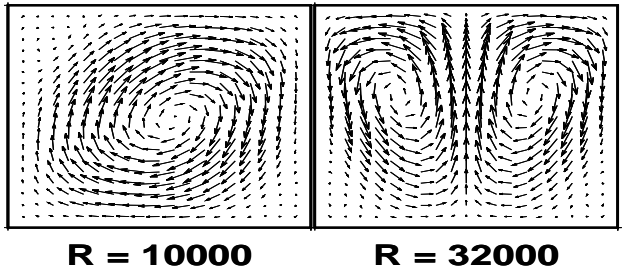}                                                                \caption{
Stationary roll patterns observed at Rayleigh Numbers of 10,000 and 32,000 with a Prandtl number of unity.  The
velocities are taken from Eulerian grid-based solutions of the conservation laws with linear Newton-Fourier
constitutive relations.
}
\end{figure}

Figure 2 shows observed stationary roll patterns typical of a Rayleigh-B\'enard flow with a gravitational
force acting in the negative y direction and a temperature gradient resulting in heat convection in the positive
$y$ direction.  The temperature and velocity are fixed on the horizontal and vertical boundaries, just as in the
idealized one- and three-roll flows of Figure 1 .  Higher values of the Rayleigh number result in solutions that
form with three or more rolls, periodic roll motions, and finally chaotic motions.  See References 6 and 7 .  

Figure 3 shows four typical {\it chaotic} Rayleigh-B\'enard velocity plots. The Rayleigh Number here is
 800 000 .  A mesh of $160 \times 160$ cells and $161 \times 161$ nodes was used.

\begin{figure}
\vspace{1 cm}
\includegraphics[height=5.5in,width=3.5in,angle= -90]{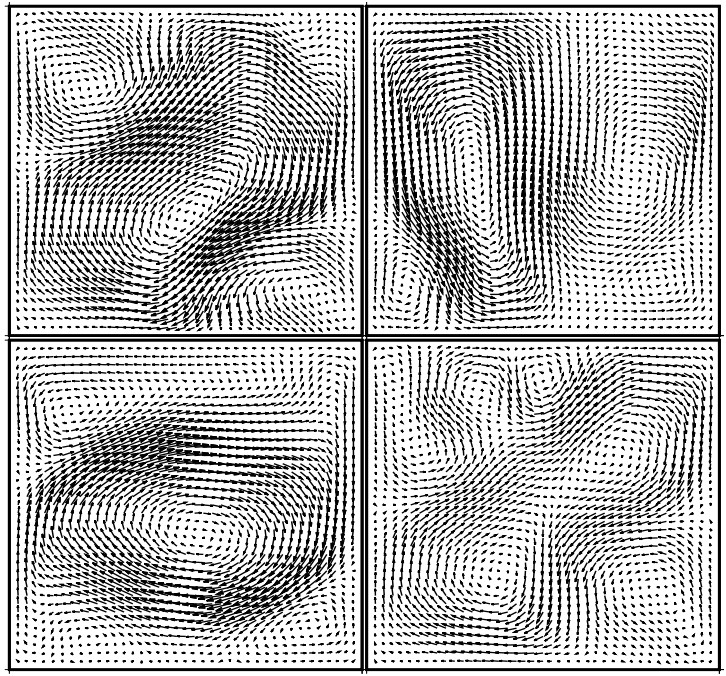}
\vspace{5 mm}
\caption{
Four snapshots of a turbulent flow computed with 160 $\times$ 160 computational cells.
}
\end{figure}

\section{Numerical Methods for Rayleigh-B\'enard Flows}

\subsection{Particle Methods: Nonequilibrium Molecular Dynamics and SPAM}

{\bf Nonequilibrium molecular dynamics} is a straightforward but limited method for studying 
Rayleigh-B\'enard flows.  Though the method is both simple and fundamental, atomistic particle studies
have several disadvantages : first, the {\it equation of state} can't be specified in advance ( only the
interatomic force law is given in molecular dynamics ) ; second, the {\it number of degrees of freedom} 
required to simulate convective rolls is either thousands ( in two dimensions\cite{b9} ) or millions ( in
three dimensions\cite{b10} ) ; third, the {\it time step} in molecular dynamics simulations is a fraction
of the collision time rather than the considerably larger time [ $dt = (dx/c)$ , where $c$ is the sound
velocity ] given by the continuum Courant condition. Finally, even with these large particle numbers and
small time steps, the
{\it fluctuations} in the atomistic simulations are so large that snapshots of nominally steady flows show
large deviations from time averages.

Mareschal and Rapaport and their coworkers\cite{b9,b10} have studied two- and three-dimensional
molecular dynamics systems, relatively large at the time they were carried out ( with 5000 and 3,507,170
particles respectively ).  In both these cases time averages were required.  The simulations did confirm
that these {\it time averages} of the atomistic flows closely matched the corresponding stationary
continuum simulations.  We have carried out a few corroborating simulations.  In these we used
thermostated boundaries composed of particles tethered to fixed lattice sites.  Rather than obeying
conservative Newtonian mechanics, here the bottom row of ``hot''  and top row of ``cold'' boundary particles
separately follow {\it thermostated} equations of motion with their temperatures controlled by the
Nos\'e-Hoover friction coefficients\cite{b7} $( \ \zeta_{\rm hot},\zeta_{\rm cold} \ )$ :
$$
\{ \ \ddot r = (F(\{ \ r \ \})/m) - \zeta_{\rm  hot} \dot r \ \}_{\rm  hot} \ ; \
\{ \ \ddot r = (F(\{ \ r \ \})/m) - \zeta_{\rm cold} \dot r \ \}_{\rm cold} \ .
$$
Particle escapes can be prevented by using a strong repulsive boundary potential to reflect
any particle venturing ``outside'' the box. Figure 4 compares a time-averaged exposure of a typical molecular dynamics run with 23,700 particles
to the final snapshot from the same simulation.

\begin{figure}
\includegraphics[width=2in,angle= -90]{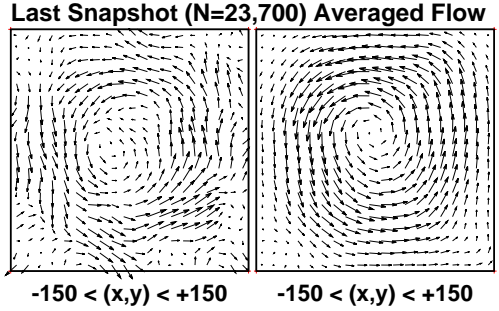}
\vspace{5 mm}
\caption{
Simulation of Rayleigh-B\'enard flow with molecular dynamics.  A snapshot, using smooth-particle
averages of the particle velocities, is at the left.  Averages appear at the right.
}
\end{figure}

{\bf Smooth Particle Applied Mechanics} ( SPAM\cite{b8} ) is a {\it macroscopic} particle alternative to molecular
dynamics.  SPAM simulations are based on the continuum constitutive relations rather than atomistic
interatomic forces.  Hundreds of particles, rather than thousands, can generate rolls, the timestep is
much larger, and individual snapshots do reproduce the stable roll structures quite well.  SPAM defines
local continuum averages by combining contributions from a few dozen nearby particles.  {\it All} of these
continuum properties, $\{ \ \rho,u,e,P,Q,\ \dots \ \}$ are {\it local averages} from sums using a weight
function like Lucy's, which is shown in Figure 5 :
$$
w(r<h) = (5/\pi h^2)(1+3z)(1-z)^3 \ ; \ z \equiv (r/h) \ .
$$
Averages computed using this twice-differentiable weight function have two continuous spatial derivatives,
enough for representing the righthand sides of the diffusive continuum equations with continuous functions.

\begin{figure}
\includegraphics[width=1.5in,angle= -90]{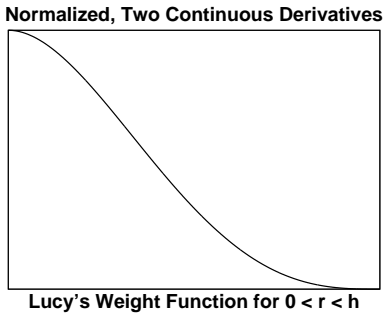}
\vspace{5 mm}
\caption{
Lucy's weight function, normalized for two dimensions, $\int_0^h2\pi rw(r<h)dr \equiv 1$ .
}
\end{figure}

Consider the simplest application of SPAM averaging, the {\it definitions} of the local densities and velocities
in terms of  smooth-particle weighted sums :
$$
\rho(r) \equiv m\sum_iw(r-r_i) \ ; \ \rho(r)u(r) \equiv m\sum_iw(r-r_i)v_i \ . 
$$
These definitions satisfy the continuum continuity equation exactly!  The variation of the density at a
fixed location $r$ can be evaluated by the chain rule :
$$
(\partial \rho/\partial t)_r = m\sum_iw'v_i \cdot (r_i-r)/|r-r_i| \ .
$$
Then notice that the gradient with respect to $r$ of the product $(\rho u)$ includes exactly the same terms,
but with the opposite signs :
$$
\nabla_r\cdot (\rho u) = m\sum_iw'v_i\cdot (r-r_i)/|r-r_i| \ .
$$
Thus {\it the SPAM version of the continuity equation} ,
$$
(\partial \rho/\partial t) \equiv -\nabla \cdot (\rho u) \longleftrightarrow \dot \rho = -\rho\nabla\cdot u \ ,
$$
{\it is an identity}, independent of the form or range $h$ of the weight function $w(r<h)$ .  This is not
entirely a surprise as there is no ambiguity in the locations of the particles' masses and momenta.

The pressure and energy are more complicated.  The smooth-particle equation of motion\cite{b8} is 
antisymmetric in the particle indices.  Thus that motion equation ,
$$
\dot v_i \equiv -m\sum_j[ \ (P/\rho^2)_i + (P/\rho^2)_j \ ] \cdot \nabla_i w(r_i-r_j) \ ;
$$
conserves {\it linear} momentum ( but {\it not} angular momentum ) exactly.  Notice that whenever the pressure
varies slowly in space the weight function plays the role of a repulsive potential with the strength of
the interparticle ``forces'' proportional to the local pressure.

The {\it gradients} in SPAM are evaluated by taking derivatives of the corresponding sums.  The temperature
gradient, for example, is :
$$
(\nabla T)_i \equiv m \sum_j(T_j-T_i)w'_{ij}[ \ (r_j-r_i)/(|r_{ij}|\rho_{ij}) \ ] \ ; \
\rho_{ij} \equiv \sqrt{\rho_i\rho_j} \ {\rm or} \ (\rho_i + \rho_j)/2 \ .
$$
Notice that two neighboring particles make no contribution to the temperature gradient if their temperatures
match.  With the gradients defined the pressure tensor and heat flux vectors can be evaluated for all the
particles and used to advance the particle properties to the next time step :
$$
\{ \ \dot r,\dot v,\dot e \ \} \ \longrightarrow \ \{ \ r,v,e \ \}
$$
In all, the SPAM method averages involve about two dozen distinct particle properties.  This computational effort
is compensated by SPAM's longer length and time scales.

In addition to providing an alternative approach to solving the continuum equations the SPAM averaging
technique can be used to average molecular dynamics properties such as $P$ and $Q$.  This approach is
particularly valuable in shockwaves ( see Chapter 6 of Reference 7 ), where constitutive properties change
on an atomistic distance scale.  It would be interesting to compare the two sides of the continuum energy
and motion equations and to use this comparison to optimize the choice of the weight function's range $h$ .

The molecular dynamics results differ qualitatively from continuum results in their time symmetry, so
that averaging offers a way of reducing this conflict.  Time-dependent solutions offer a specially flexible
technique for bringing the two approaches into better agreement.  The weight functions also offer a way
of carrying out the coarse graining which could be used to reduce the conflict between the microscopic and
macroscopic forms of mechanics.

\subsection{Eulerian Finite-Difference Method\cite{b9}}

Straightforward centered-difference approximations to the continuum equations provide a useful approach
to the Rayleigh-B\'enard problem.  Mareschal and his coworkers pointed out that an efficient numerical
method can be based on square cells or zones, with the velocities and energies defined at the nodes and
the densities defined in the cells.  A small $10 \times 10$ cell program written in this way would solve
$3\times 11 \times 11 + 10 \times 10 = 463$ ordinary differential equations.  The solution procedure
follows a seven-step plan: [1] use linear interpolation and extrapolation near the boundaries to find
the complete set of 484 nodal variables and 400 cell variables; [2] use centered differences to find
$\nabla u$ and $\nabla T$ ; [3] use these gradients to obtain $P$ and $Q$ ; [4] evaluate $\nabla u:P$ and
$\nabla \cdot Q$ ; [5] evaluate $(\partial \rho/\partial t)$ from the neighboring nodal values ; [6]
evaluate $(\partial u/\partial t)$ from the pressure gradients and $(\partial e/\partial t)$ by summing
the convective contributions and the work and heat; [7] use fourth-order Runge-Kutta integration to advance
the 463 dependent variables to the next time step.

The Rayleigh-B\'enard solutions -- simple rolls\cite{b11}, periodic solutions, or chaos -- can be observed
in either two dimensions, where there are plenty of puzzles to solve, or three.  Because simulation and
visualization are simpler in two dimensions, while the challenges to understanding remain severe, we choose
two dimensions.  The critical Rayleigh Number of about 5000 corresponds to an eddy width which can easily
be resolved with $8 \times 8$ cells and a one-roll Reynolds number of order unity.

The Rayleigh number varies as $L^4$.  Doubling the sidelength $L \rightarrow 2L$ with $g$ and the transport
coefficients fixed changes ${\cal R}:5000 \rightarrow 80000$ and increases the number of rolls to four.
Desktop or laptop machines are quite capable of simulations with $ {\cal R} = 1 \ 000 \ 000$ , for which
this simple-minded reasoning could lead us to expect about $(1000000/5000)^{1/4} \simeq 14$ rolls.  In
fact this doesn't happen.  See Figure 7 below. In two dimensions the energy flow is {\it toward}, rather
than away from,
large rolls.  For ${\cal R} = 800K$ [ $K$ indicates thousands ] and a $160 \times 160$ mesh one finds
occcasional deep minima in the time-dependent kinetic energy.  These minima correspond to only two large rolls,
as in the simple solutions without chaos, with ${\cal R} \simeq 10K$ .  In three dimensions the chaotic flow
is qualitatively different, and more complicated.  Instead of whirling vortices one finds plumes ascending
and descending, with mushroom shaped heads for large Prandtl numbers ( glycerin ) where viscosity dominates
conductivity\cite{b1}.

\begin{figure}
\vspace{1 cm}
\includegraphics[width=2in,angle= -90]{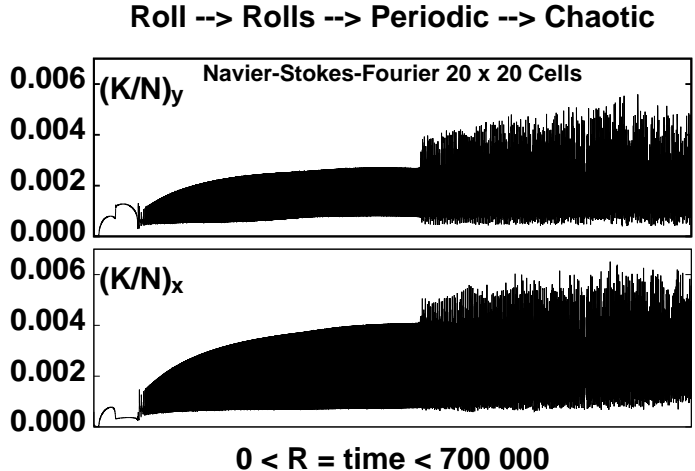}
\vspace{5 mm}
\caption{
Kinetic energy per cell (vertical at top and horizontal at bottom) from a simulation with ${\cal R} = t$.
Notice, at the extreme left of the two plots, the reduction in the horizontal kinetic energy at the transition
from one roll to two ( ${\cal R} \simeq 25K$ ) .
}
\end{figure}
Figure 6 shows the variation of the kinetic energy per cell with the Rayleigh Number, where
${\cal R} = t \leq 700K$.
The transitions go from one roll to two, and from two rolls to a time-periodic arrangement with perhaps
four, which in turn gives rise to chaos.  Our simple centered-finite-difference fixed-timestep code
``blew up'' at Rayleigh numbers of
$$
\{ \ 715K, \ 810K, \ 840K, \ 860K, \ 905K, \ 940K \ \} \ {\rm for} \ 
L = \{ \ 16, \ 24, \ 32, \ 48, \ 64, \ 96 \ \} \ .
$$
Figure 6 suggests that chaos sets in around
${\cal R} = 385K$ and gradually increases in strength until the algorithm becomes unstable for the
chosen mesh.

The motion responds relatively quickly to perturbations.  To demonstrate this we show in Figure 7 six
snapshots from a $24 \times 24$ simulation with {\it all} of the flow velocities
{\it instantaneously} reversed from forward to backward at time 0, where the forward chaotic
flow has three distinct rolls.  By a time of 500 (a few dozen sound traversal times) the flow is
back to normal, again with three rolls.  But in the transformational process of discarding the
time-reversed morphology the stabilizing flow acquires as many as seven rolls, in accord with the
estimate that roughly $8 \times 8$ cells are required to resolve a roll.  The quick reduction
in roll number as steady chaos is regained is again a symptom of the flow of energy from smaller to
larger rolls in two dimensions.

\begin{figure}
\vspace{1 cm}
\includegraphics[width=3in,angle= -90]{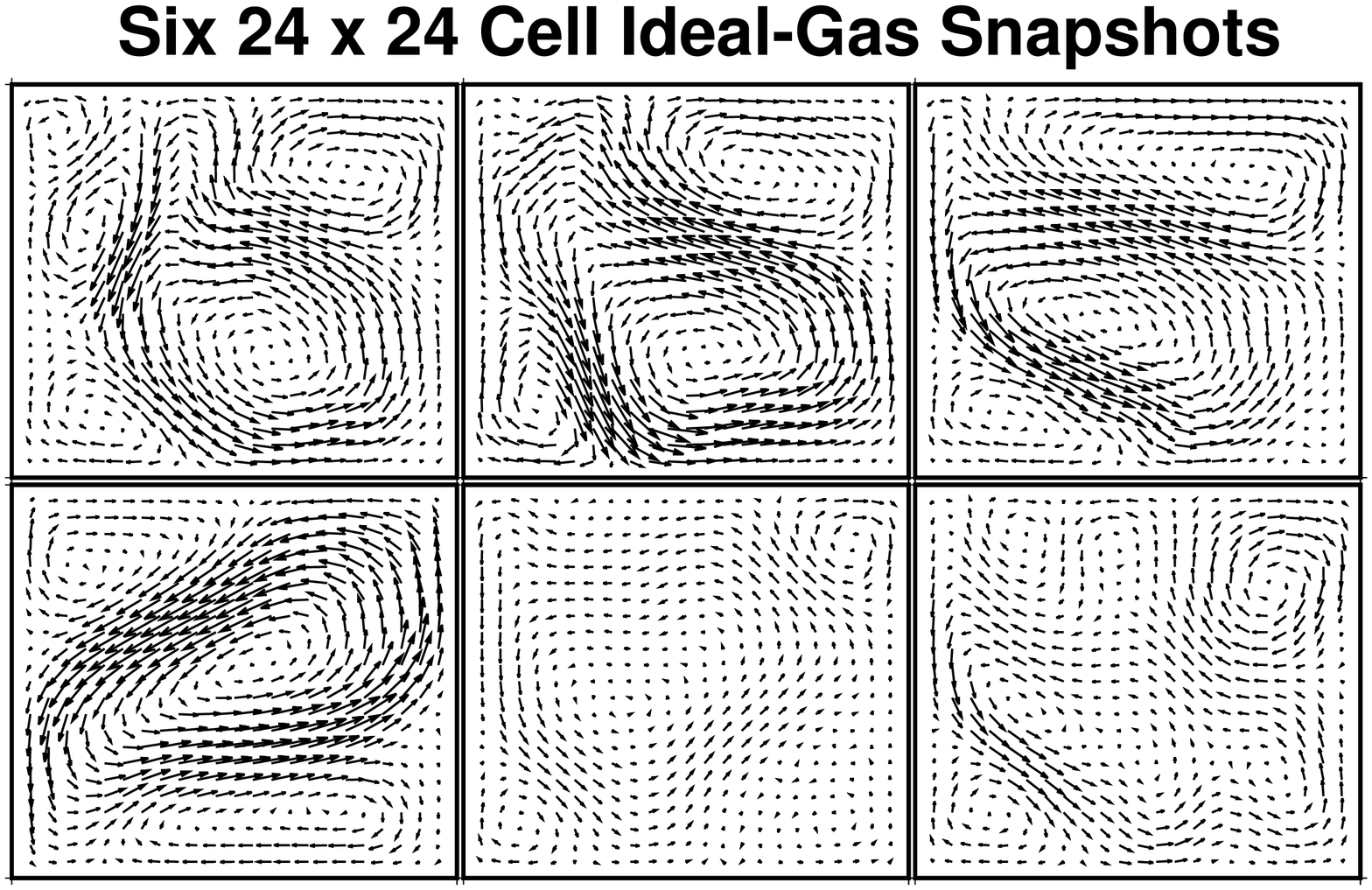}
\vspace{5 mm}
\caption{
Snapshots at time 0, 100, 200 (bottom row, left to right) and 300, 400, 500 (top row, left to right)
starting with an instantaneously reversed chaotic flow with ${\cal R} = 800 \ 000$ .
}
\end{figure}

Although there is no difficulty in simulating the motion on larger meshes even the modest 576-cell
simulation of Figure 7 provides an excellent characterization of chaos.  Characterizing chaos quantitatively
entails evaluating Lyapunov exponents, the tendency of nearby trajectories to separate farther or to
approach one another.  The separation of a satellite trajectory from its reference has four component types :
$$
\delta \equiv ( \ \{ \delta \rho \} , \{ \delta u_x \} , \{ \delta u_y \} , \{ \delta e \} \ ) \ .
$$
Keeping the separation constant by rescaling at every timestep gives the local exponent ,
$$
\lambda_i(t) = \ln(| \ \delta_{\rm before} \ |/| \ \delta_{\rm after} \ |)/dt \ .
$$

\begin{figure}
\vspace{1 cm}
\includegraphics[width=2in,angle= -90]{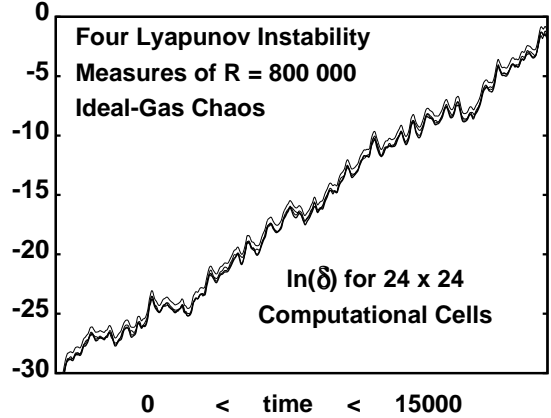}
\vspace{5 mm}
\caption{
Separation between two strongly chaotic simulations.  The fourth-order Runge-Kutta timestep is 0.05.
Although four different sets of data (for $\rho, \ u, \ e$ subspaces as well as the complete space)
are shown, here, the local Lyapunov exponents are scarcely distinguishable from one another.
}
\end{figure}

In atomistic simulations it is well known that the largest Lyapunov exponent can be measured in
configuration, momentum, or phase space, with identical longtime averages\cite{b12}.  In continuum
simulations one would expect that the density, velocity, or energy subspaces could be used in this
same way.  By choosing particular mass, length, and time units any one of the three subspaces could
be made to dominate the local Lyapunov exponent.

Figure 8 shows the exponential separation of a satellite trajectory from its reference trajectory,
without rescaling, as measured in the density, velocity, energy, and complete state spaces of the flow.
The similarity of the four curves is so complete that we don't attempt to label them separatedly in
the figure.  Evidently, despite the changing morphology of the chaotic vortices, the underlying chaos
( at a Rayleigh Number of 800$K$ ) is quite steady.  The research literature indicates that the Lyapunov
spectrum in such two-dimensional flows is roughly linear, and corresponds to a strange attractor with
only a few degrees of freedom, no doubt corresponding to the number of observed vortices.

The low dimensionality of two-dimensional turbulent chaos results from the tremendous dissipation
inherent in the Navier-Stokes-Newton-Fourier model.  If we evaluate the three ``phase-space derivatives''
which contribute to the continuum analog of Liouville's particle Theorem :
$$
(\partial \dot \rho/\partial \rho) \ ; \ (\partial \dot u/\partial u) \ ; \ (\partial \dot e/\partial e) \ ,
$$
[ here the dots are time derivatives {\it at fixed cells or nodes} ] the velocity and energy derivatives
give $-4(\eta/\rho)/(dx)^2$ and $-4\kappa/(dx)^2$ respectively while the density derivative vanishes.  In
the end only a few degrees of freedom exhibit chaos.

\section{Conclusions and Suggestions for Research}

Continuum mechanics can be studied with finite-difference ordinary differential equations, or with particle
differential equations, resembling those used in molecular dynamics.  The finite-difference approach is
certainly the most efficient of these possibilities.  The Rayleigh-B\'enard problems exhibit a variety of
flows, with interesting results at the level of a few hundred compuational cells.  The relative stability
of the flows and the characterization of the chaos are both interesting research areas.  With the limited
dimensionality of the chaotic flows' attractors, estimating only a few Lyapunov exponents suffices to
characterize Rayleigh-B\'enard chaos.  The loci of Lyapunov vectors' instability is yet another source of
fascinating questions and answers.

Though we have no space to discuss them here the useful smooth-particle technique for bridging
together the particle and continuum methods suggests a variety of problems designed to reduce the discrepancies
between the three types of numerical algorithm.

\section{Acknowledgments} We first learned about the Rayleigh-B\'enard problem through a lecture by Jerry
Gollub at Sitges some 25 years ago.  That talk led to Ph D projects for Oyeon Kum and Vic Castillo at the
University of California's Davis/Livermore Campus' ``Teller Tech''.  Their work continues to provide us
with inspiration.  We thank Denis Rapaport for helpful emails and
Anton Krivtsov and Vitaly Kuzkin for encouraging our preparation of this lecture.


\newpage
\begin{thebibliography}{99}


\bibitem{b1} L. P. Kadanoff, ``Turbulent Heat Flow: Structures and Scaling'', Physics Today
34-39 (August 2001).

\bibitem{b2} N. T. Ouellette, ``Turbulence in Two Dimensions'', Physics Today 68-69 (May 2012).

\bibitem{b3} G. E. Karniadakis and S. A. Orszag, ``Nodes, Modes, and Flow Codes'', Physics
Today 34-42 (March 1993).

\bibitem{b4} A. E. Deane and L. Sirovich, ``A Computational Study of Rayleigh-B\'enard
Convection. Part 1. Rayleigh-Number Scaling [ and ] Part 2. Dimension Considerations'',
Journal of Fluid Mechanics {\bf 222}, 231-250 [ and ] 251-265 (1991).

\bibitem{b5} R. H. Kraichnan and D. Montgomery, ``Two-Dimensional Turbulence'', Reports on
Progress in Physics {\bf 43}, 547-619 (1980).

\bibitem{b6} A. V. Getling, {\it Rayleigh-B\'enard Convection -- Structures and Dynamics}.
(World Scientific, Singapore, 1998).

\bibitem{b7}  Wm. G. Hoover and Carol G. Hoover, {\em Time Reversibility, Computer Simulation,                          
Algorithms, and Chaos} (Second Edition, World Scientific, Singapore, 2012) .

\bibitem{b8} Wm. G. Hoover, {\it Smooth Particle Applied Mechanics: the State of the Art},
World Scientific, Singapore, 2006) .

\bibitem{b9} A. Puhl, M. M. Mansour, and M. Mareschal, ``Quantitative Comparison
of Molecular Dynamics with Hydrodynamics in Rayleigh-B\'enard Convection'',
Physical Review A {\bf 40}, 1999-2012 (1989).

\bibitem{b10} D. C. Rapaport, ``Hexagonal Convection Patterns in Atomistically Simulated
Fluids'', Physical Review E {\bf 73}, 025301R (2006).

\bibitem{b11} V. M. Castillo, Wm. G. Hoover, and C. G. Hoover, ``Coexisting Attractors in
Compressible Rayleigh-B\'enard Flow'', Physical Review E {\bf 55}, 5546-5550 (1997).

\bibitem{b12} S. D. Stoddard and J. Ford, ``Numerical Experiments on the Stochastic Behavior
of a Lennard-Jones Gas System'', Physical Review A {\bf 8}, 1504-1512 (1973).

\end{thebibliography}
\end{document}